\documentclass[a4paper,11pt]{article}
\usepackage{pos}

\def\be{\begin{eqnarray*}}
\def\ee{\end{eqnarray*}}

\def\bea{\begin{eqnarray}}
\def\eea{\end{eqnarray}}

\begin{document}
\title{Hadronic vacuum polarization of the muon on 2+1+1-flavor HISQ ensembles: an update}
\author*[a]{Shaun Lahert}
\author[b]{Carleton DeTar}
\author[a]{Aida X.~El-Khadra}
\author[c]{Elvira G{\'a}miz}
\author[d]{Steven Gottlieb}
\author[e]{Andreas S.~Kronfeld}
\author[f]{Ethan T.~Neil}
\author[f]{Curtis T.~Peterson}
\author[e]{Ruth Van de Water}


\affiliation[a]{Department of Physics, University of Illinois Urbana-Champaign, Urbana, IL, USA}
\affiliation[b]{Department of Physics and Astronomy, University of Utah, Salt Lake City, UT, USA}
\affiliation[c]{Granada U., Theor. Phys. Astrophys., Campus de Fuentenueva, Granada, Spain}
\affiliation[d]{Department of Physics, Indiana University, Bloomington, IN, USA}
\affiliation[e]{Particle Physics Department, Theory Division, Fermilab, Batavia, IL, USA}
\affiliation[f]{Department of Physics, University of Colorado Boulder, Boulder, CO, USA}

\emailAdd{slahert@illinois.edu}
\abstract{
\textbf{\textsf{Fermilab Lattice, HPQCD, and MILC Collaborations}}\vskip1em
We give an update on the status of the Fermilab Lattice-HPQCD-MILC calculation of the contribution to the muon's anomolous magnetic moment from the light-quark, connected hadronic vacuum polarization. We present preliminary, blinded results in the intermediate window for this contribution, $a_{\mu, \textrm{W}}^{ll}$. The calculation is performed on $N_f =2+1+1$ highly-improved staggered quark (HISQ) ensembles from the MILC collaboration with physical pion mass at four lattice spacings between 0.15~fm and 0.06~fm. We also present preliminary results for a study of the two-pion contributions to the vector-current correlation function performed on the 0.15~fm ensemble where we see a factor of four improvement over traditional noise reduction techniques.}

\FullConference{The 38th International Symposium on Lattice Field Theory - LATTICE21\\
ZOOM/GATHER @ MIT\\
26–30 Jul 2021
}
\maketitle

\section{Introduction}

\vspace{-3mm}

In April 2021, the FNAL E989  experiment released their first measurement of the muon's anomalous magnetic moment \cite{PhysRevLett.126.141801}. Combined with the previous result from the BNL E821 experiment \cite{Bennett:2006fi} the disagreement with the Standard Model prediction of Ref.~\cite{Aoyama:2020ynm} stands at 4.2$\sigma$. The error on the experimental average is now  0.35 ppm, and the Fermilab experiment plans to increase the precision by a factor of four. The Standard Model prediction of Ref.~\cite{Aoyama:2020ynm} quotes an uncertainty of 0.37 ppm, dominated by the leading-order hadronic vacuum polarization (HVP).  
The most precise theoretical determinations of the hadronic vacuum-polarization contribution come from a data-driven, dispersive approach using experimental data for the low-energy hadronic $e^+e^-$ cross section \cite{Aoyama:2020ynm, PhysRev.168.1620} as inputs.\footnote{See Ref.~\cite{Aoyama:2020ynm} for complete references to the experimental and theoretical papers leading to the data driven value of the HVP contribution to the muon anomalous magnetic moment.}
Lattice QCD provides an alternative {\em ab initio} approach that is independent of these experimental inputs and where all sources of uncertainty can be controlled and systematically reduced. Recently, a lattice QCD calculation \cite{Borsanyi:2020mff} obtained a result with sub-percent precision, much closer to the experimental average (1.5$\sigma$),  in 2.1$\sigma$ tension with the SM prediction of Ref.~\cite{Aoyama:2020ynm}.  As such, it is of utmost importance to confront the calculation in Ref.~\cite{Borsanyi:2020mff} with other, independent lattice QCD calculations of commensurate precision.

In this proceedings, we give an update on the ongoing effort, of the Fermilab Lattice, HPQCD, and MILC collaborations, to compute the leading-order HVP contribution to the muon's anomalous magnetic moment, $a_{\mu}^{\mathrm{HVP}, \mathrm{LO}}$. In particular, here we focus on the dominant light-quark, connected contribution ($a_{\mu}^{ll}$), we refer readers to Ref.~\cite{cmcneile} for an update of our disconnected and QED + QCD calculations. We present preliminary, blinded results for the intermediate window $a_{\mu, \textrm{W}}^{ll}$ \cite{RBC:2018dos}. This sub-quantity is obtained by using a smooth `window' function to restrict, to an intermediate region, the HVP integral over Euclidean time of the light-quark, connected, vector-current two-point correlation function. This quantity is important as it can be computed to high precision with significantly less effort than the full $a_{\mu}^{ll}$ and is less dependent on lattice effects. 
This calculation is done on the MILC $N_f=2+1+1$ highly-improved staggered quark (HISQ) ensembles \cite{Bazavov:2012xda} with physical pion mass at four lattice spacings between 0.15~fm and 0.06~fm. 

As mentioned, precise lattice determinations of the full $a_{\mu}^{ll}$ present many technical and computational challenges \cite{Davies:2019efs}, notably, the significant statistical noise from the large-time region of the vector-current correlation function. We present preliminary results from a new study to determine, exactly, the two-pion contributions to the HVP that dominate this region. Performed on our coarsest ensemble with $a\approx 0.15$~fm, using explicit staggered two-pion operators to resolve the lowest-energy states, we reconstruct the light-quark, connected, vector current two-point function. This approach results in an improvement over traditional noise-reduction techniques \cite{RBC:2018dos} by a factor of~four.

\vspace{-1mm}

\section{Light-quark, connected HVP}



In the standard model, the HVP in Euclidean space is given by 
\bea
\Pi^{\mu v}(q^{2})&=&\left(\delta^{\mu v} q^{2}-q^{\mu} q^{v}\right) \Pi(q^{2})=\int d^{4} x\, e^{i q x}\left\langle J^{\mu}(x) J^{v}(0)\right\rangle\\
J^{\mu}(x)&=&\sum_{i} Q_{i} \bar{\psi}_{i}(x) \gamma^{\mu} \psi_{i}(x)\, . \label{vecCurrent}
\eea
Here, $J^{\mu}(x)$ is the electromagnetic vector current for the quarks, and $Q_i$ are the corresponding charges in units of electron charge. The leading-order contribution to the muon's anomalous magnetic moment from the HVP is \cite{Blum:2013xva} 
\bea
a_{\mu}^{\mathrm{HVP}, \mathrm{LO}}&=&4\alpha^{2} \int_{0}^{\infty} d Q^{2} K_{E}(Q^{2}) \widehat{\Pi}(Q^{2})\, ,
\eea
where $\hat{\Pi}\left(q^{2}\right)=\Pi\left(q^{2}\right)-\Pi(0)$ is the subtracted vacuum polarization. The integration kernel $K_{E}(Q^{2})$ depends on the muon's mass. To compute $a_{\mu}^{\mathrm{HVP}, \mathrm{LO}}$ on the lattice, it is convenient to use the time-momentum representation introduced in \cite{Bernecker:2011gh}, namely,
\bea
a_{\mu}^{\mathrm{HVP}, \mathrm{LO}}&=&4\alpha^{2} \int_{0}^{\infty} d t\, \tilde K(t) C(t)  \label{amuTint}\\
\tilde{K}(t) &\equiv& 2 \int_{0}^{\infty} \frac{\mathrm{d} Q}{Q} K_{E}(Q^{2})\left[Q^{2} t^{2}-4 \sin ^{2}\left(\frac{Q t}{2}\right)\right] \\
C(t)&=&\frac{1}{3} \sum_{\vec{x}, k}\left\langle J^{k}(\vec{x}, t) J^{k}(0)\right\rangle \label{corrFunc2pt}
\eea
with $k=1,2,3$. The two-point vector-current correlation function $C(t)$ receives contributions from all quark flavors $u,d,s,c,t,b$ and connected and disconnected Wick contractions. We work in the isospin-symmetric limit $u=d=l$ and only consider the dominant light-quark, connected contribution, $a_{\mu}^{ll}$, here.
\begin{figure}[t]
  \centering
    \includegraphics[width=.87\textwidth]{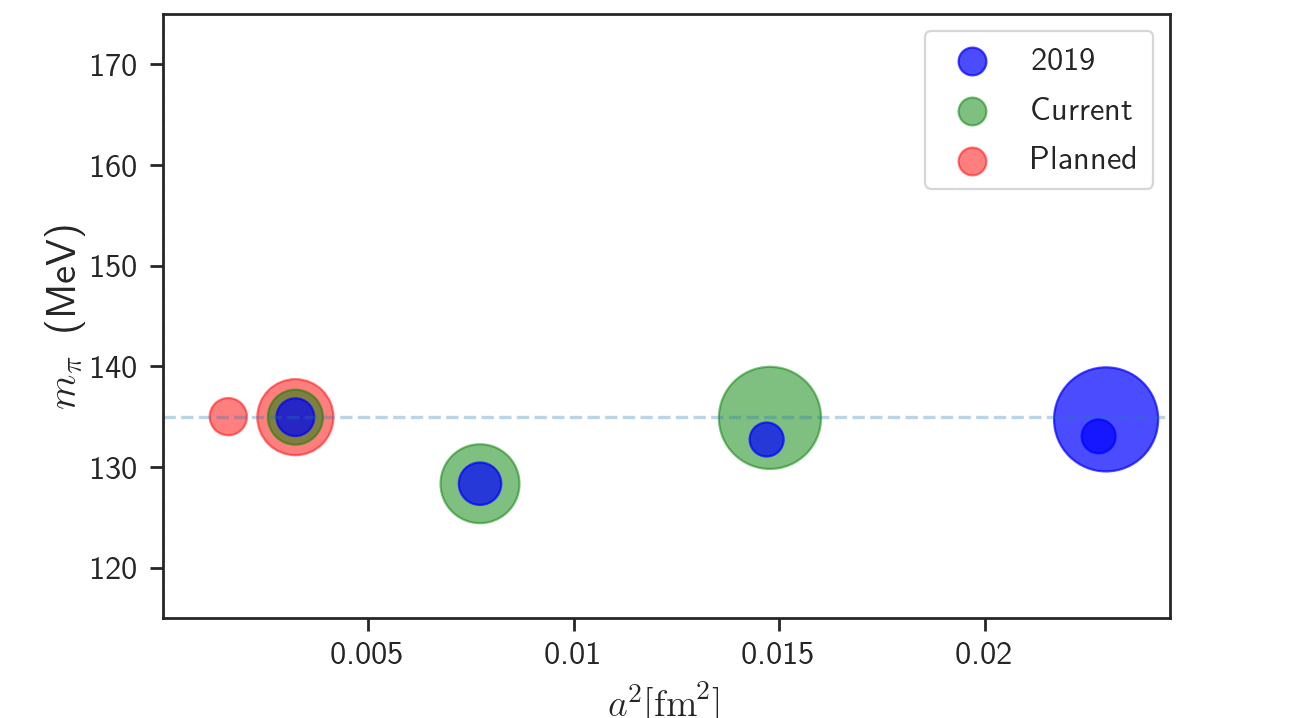}
    \caption{Plot showing the HISQ ensemble parameters, pion mass $m_\pi$ and lattice spacing $a^2$ with number of configs indicated by area of the circles, used in Ref.~\cite{Davies:2019efs} (blue), current statistics (green), planned (red).}
    \label{fig:ens}
\end{figure}
Some parameters of the HISQ ensembles used in this work are shown in Fig.~\ref{fig:ens}; most of our ensembles have close to physical pion masses and have box size  $L > 3.7/m_\pi$. 
As described in Ref.~\cite{Davies:2019efs}, the correlation function in Eq.~(\ref{corrFunc2pt}) is computed using the truncated-solver method (TSM) \cite{TSM1, TSM2} with a random-wall source, with one fine-residual conjugate gradient solve and a number of loose-residual solves, as shown in Table~\ref{table:enssolves}. 
\begin{table}[h]
\centering
\begin{tabular}{l|c|c|c|c}
$\mathrm{\approx a}[\mathrm{fm}]$ & $N_s^3 \times N_t$ & $a m_{l}^{\text {sea }} / a m_{s}^{\text {sea }} / a m_{c}^{\text {sea }}$ &$N_{\text{cfg}}$ & $N_{\text {src}}$ \\ 
\hline
$0.15$ & $ 32^3 \times 48$ & 0.002426 / 0.0673 / 0.8447 & 9362 & 48 \\ 
$0.12$ & $ 48^3 \times 64$ & 0.001907 / 0.05252 / 0.6382 & 9011 & 64 \\ 
$0.09$ & $64^3 \times 96$ & 0.00120 / 0.0363 / 0.432 & 5384 & 48 \\ 
$0.06$ & $96^3 \times 128$ & 0.0008 / 0.022 / 0.260 & 2621 & 24 \\ 
\end{tabular}
\caption{Ensemble parameters used in this work. Column 2 gives the volume of the lattices in number of space-time points. Column 3 gives the number of configurations analyzed.  Column 4 gives the number of loose-residual solves performed on each configuration for use in the truncated solver method \cite{TSM1, TSM2}. 
}
\label{table:enssolves}
\end{table} 

Reference~\cite{Davies:2019efs} gave a value of $637.8(8.8)$ for the light-quark, connected contribution to the muon's anomalous magnetic moment. In order to reach sub-percent precision, we have significantly increased our statistics on the three finest ensembles, as indicated in Fig.~\ref{fig:ens}. We are currently testing exact low-eigenmode approaches for computing the vector current propagator to improve our large-time determination of the correlation function. Aside from statistics, the largest sources of error were the continuum extrapolation uncertainty, (finite volume, staggered taste-breaking, $m_\pi$), and scale setting (absolute). To improve the continuum extrapolation, we are increasing our statistics on the $\approx 0.06$~fm ensemble and are beginning efforts to add data on a $\approx 0.042$~fm ensemble Fig.~\ref{fig:ens}. Ref.~\cite{Davies:2019efs} used the chiral model of Ref.~\cite{JegerlehnerModel, Chakraborty:2017tqp} to perform all lattice corrections.  We are now developing independent analysis strategies for each correction.
There is also an ongoing effort to address the absolute scale setting uncertainty on HISQ ensembles, using an improved determination of the $\Omega$ baryon mass on all ensembles. \cite{Chuges}.


\subsection{Intermediate window}


Precise lattice determinations of $a_{\mu}^{ll}$ are hindered by large statistical noise in the long Euclidean time region, significant finite volume effects (even on ensembles with  $m_\pi L > 4$) and short-range discretization effects \cite{Chakraborty:2017tqp, Davies:2019efs}. As a first step, it makes sense to compute a sub-quantity which 
minimizes or isolates these effects. The $a_{\mu}^{\mathrm{HVP}, \mathrm{LO}}$ windows were introduced to achieve this by using a smoothly varying window function to restrict the region of Euclidean time over which the vector current correlation function is integrated \cite{RBC:2018dos}.
\bea
a^{\mathrm{HVP}, \mathrm{LO}}_{\mu, \textrm{win}} &=& 4 \alpha^{2} \int_{0}^{\infty} \mathrm{d} t C(t) \tilde{K}(t) \Theta\left(t, t_1, t_2, \Delta\right)\\
\Theta\left(t, t_1, t_2, \Delta\right)&=&\frac{1}{2}\left[\tanh \left(\frac{t-t_{1}}{\Delta}\right)-\tanh \left(\frac{t-t_{2}}{\Delta}\right)\right] \, .
\eea
Here, $t_1,\, t_2$ control the location of the window boundary, and $\Delta$ controls sharpness. A value of $\Delta$ is chosen to be as large as the largest lattice spacing considered, typically $0.15$~fm. The intermediate  window `$\mathrm{~W}$' with parameters $[t_1, t_2]=[0.4,1]$~fm and $\Delta=0.15$~fm provides a quantity $a_{\mu, \textrm{W}}^{ll}$ which is mostly independent of the long range statistical noise, finite volume, and staggered taste-breaking effects of the low-energy two-pion contributions. It is also free of the strongest short-range discretization effects.

\begin{figure}[t]
  \centering
    \includegraphics[width=.84\textwidth]{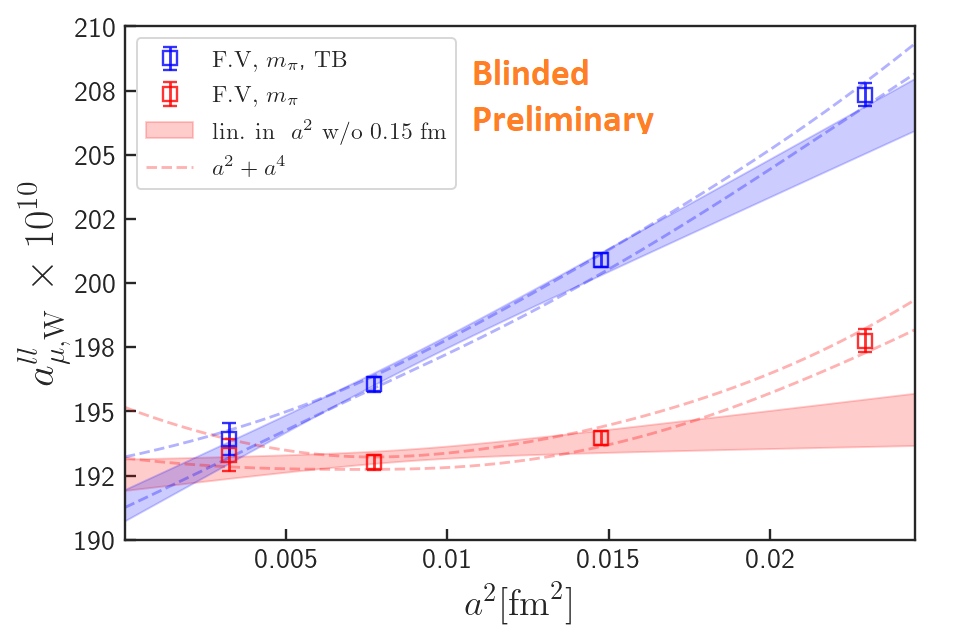}
    \caption{$a_{\mu, \textrm{W}}^{ll}$ vs lattice spacing ($a^2$) on four ensembles. Two correction schemes, finite volume + $m_\pi$ (red), finite volume + $m_\pi$ + taste breaking (blue). Solid bands (dotted lines) are linear (quadratic) extrapolations in $a^2$.}
    \label{fig:win}
\end{figure}

Our calculation for the intermediate window, $a_{\mu, \textrm{W}}^{ll}$ on the four ensembles is shown in Fig.~\ref{fig:win}. The red points have small finite volume and $m_\pi$-mistuning corrections added to the raw data, from the leading order term in the chiral model, see Table \ref{table:corrections}. The blue points include taste-breaking corrections; we ascribe no additional error for these corrections as in \cite{Chakraborty:2017tqp, Davies:2019efs}, instead taking the spread of the extrapolations as an error. In Fig.~\ref{fig:win},  we include linear extrapolations, dropping our coarsest ensemble, and extrapolations including an $a^4$ term. We observe, without correcting for taste-breaking, our data already have small discretization effects. Correcting for taste-breaking reduces the $a^4$ dependence while introducing a more prominent $a^2$ dependence. We find reasonable agreement in all continuum extrapolations with good fit quality ($\chi^2 / \textrm{d.o.f} \sim 1$), especially for the linear fits.  We plan to include further variations on these fit functions.

\begin{table}[h]
\centering
\begin{tabular}{c|c|c}
$\mathrm{\approx a}[\mathrm{fm}]$ & $m_{\pi}$ & $\mathrm{FV}$  \\
\hline 
$0.15$ & $-0.01$ \% & $0.57$ \%   \\
$0.12$ & $-0.004$ \% & $0.26$ \%    \\
$0.09$ & $-0.305$ \%& $0.35$ \%   \\
$0.06$ & $-0.001$ \% & $0.42$ \%    \\
\end{tabular}
\caption{Lattice corrections added to $a_{\mu, \textrm{W}}^{ll}$ using the leading order term in the chiral model of \cite{JegerlehnerModel, Chakraborty:2017tqp}}
\label{table:corrections}
\end{table}

\vspace{-1.5mm}
\subsection{Two-pion contribution}

\begin{figure}[t]
  \centering
    \includegraphics[width=0.8\textwidth]{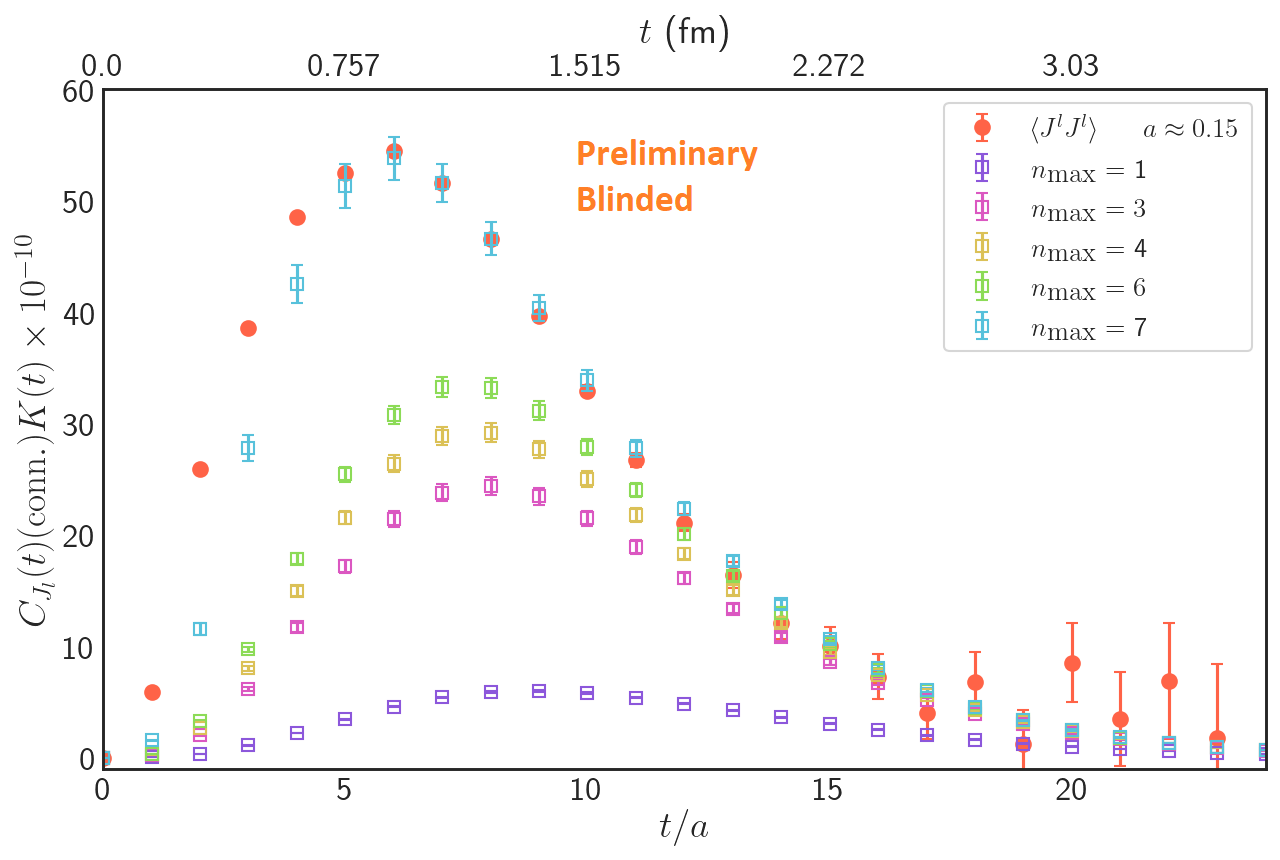}
    \caption{Blinded results for taste singlet vector current two-point, correlation function (orange). Blinded reconstruction of the correlation function from determined parameters for states up to $n_\textrm{max}=1-7$.}
    \label{fig:recon}
\end{figure}

As mentioned, a significant challenge in lattice determinations of $a_{\mu}^{ll}$ is the statistical noise in the tail of the light-quark, connected component. This can be traced to the fact that the variance of the light-quark component of the vector current, two-point function Eq.~(\ref{corrFunc}) falls off with an exponent of $ m_\pi$ \cite{Lepage:1989hd} while the signal falls of with lowest energy two-pion state, $\sim 2m_\pi $. Hence,  the noise overwhelms the two-pion contribution in the large-time region,  Fig.~\ref{fig:recon} (orange points).
\bea
C_{J_l}(t) &=& \frac{1}{3} \sum_{\vec{x}, k}\left\langle J_l^{k}(\vec{x}, t) J_l^{k}(0)\right\rangle =  \sum_n \langle 0 | J_l | n \rangle e^{-E_n t} \, . \label{corrFunc}
\eea
A remedy is to employ explicit two-pion operators to resolve the lowest lying states of the theory and reconstruct the correlation function after some Euclidean time with this information  \cite{bruno2019exclusive, PhysRevD.101.054504}. The computational strategy is then to construct a correlation-function matrix of the form
\bea
\mathbf{C}(t)=\left(\begin{array}{ll}
C_{J_l, \tilde{J}_l \rightarrow J_l, \tilde{J}_l}(t) & C_{J_l, \tilde{J}_l \rightarrow \pi \pi}(t) \\
C_{\pi \pi \rightarrow J_l, \tilde{J}_l}(t) & C_{\pi \pi \rightarrow \pi \pi}(t)
\end{array}\right) \label{ctmat}
\eea

\noindent where we have chosen the taste-singlet vector current operator (smeared) $J_l (\tilde J_l)$ as opposed to the taste-vector vector current, used in the work described above, as it has the preferable lowest-energy two-pion state, namely, two Goldstone-boson pions. The staggered two-pion operators are constructed from linear combinations of products of staggered single-pion operators which transform in the same taste representation as the vector current. The staggered single- \& two-pion operators are defined as.
\bea
\pi^{-}(\vec{p})_{\xi}&\equiv&\sum_{x} e^{i \vec{p} \cdot \vec{x}} \bar{u}(x) \gamma_{5} \otimes \gamma_{\xi} d(x)\\
\mathcal{O}_{\pi \pi}(0)&\equiv&\sum_{\xi_{1}, \xi_{2} \atop p \in\{p\}^{\star}} C G_{\text {stag, iso. }}(\xi_{1}, \xi_{2}, \vec{p}) \pi(\vec{p})_{\xi_{1}} \pi(-\vec{p})_{\xi_{2}}
\eea
where $p^\star$ is the momentum orbit. $C G_{\text {stag, iso. }}(\xi_{1}, \xi_{2}, \vec{p})$ are the Clebsch-Gordon coefficients of the isospin-staggered group \cite{Kilcup:1986dg} defined below.
\bea
S U_{I}(2) \times\left(T_{N}^{3} \rtimes\left\{\Xi_{\mu}, C_{0}\right\} \rtimes\left\{\tilde{R}_{i j}, I_{S}\right\}\right)=S U(2) \times\left(Z_{N}^{3} \rtimes \Gamma_{4,1} \rtimes W_{3}\right)
\eea
The Clebsch-Gordon coefficients are obtained by explicit construction of the irreducible representations (irreps) of the group by a double application of Wigner's method for semi-direct product groups \cite{Kilcup:1986dg}. With the irreps in hand one can extract the coefficients through the well established methods for discrete groups \cite{Sakata:1974hd, jlab}.
We choose a range of back-to-back pion momenta up to the energy at which the free two-pion state equals the mass of the rho resonance. We include all rows of the multi-dimensional taste-irrep pions. At non-zero momentum these three dimensional irreps split into non-degenerate irreps which must be accounted for. For the $\approx0.15$~fm ensemble, the range of momentum and taste is shown in Table~\ref{table:states}. 
Eigenvectors $v_n$ are extracted from a generalized eigenvalue problem (GEVP) \cite{Blossier:2009kd} applied to,
\bea
\mathbf{C}(t) v=\lambda \mathbf{C}\left(t_{0}\right) v .
\eea
\begin{figure}[t]
  \centering
    \includegraphics[width=0.93\textwidth]{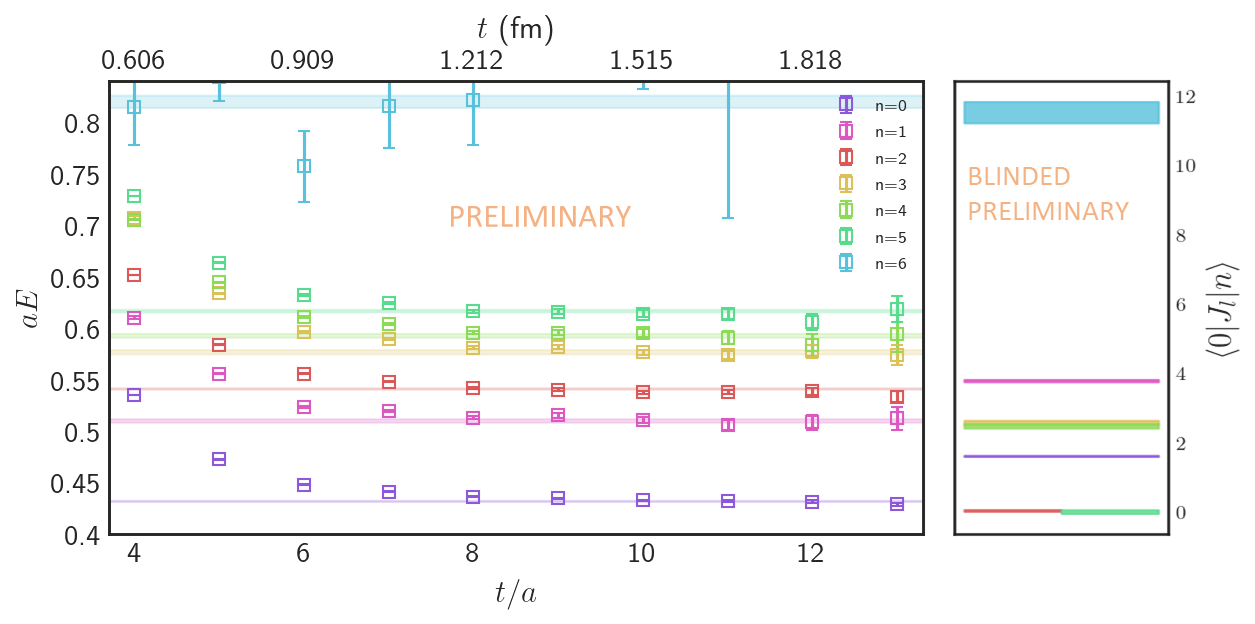}
    \caption{Energies (left) and amplitudes (right) for the two-pion states extracted from a generalized eigenvalue analysis on the $\approx 0.15$~fm ensemble. Bands are results from fits, points are effective masses.}
    \label{fig:gevp}
\end{figure}
\begin{table}[t]
\centering
\begin{tabular}{l|l}
Operator & Momentum (back-to-back) \\ \hline
$J_l, \tilde J_l$ & \\
$\mathcal{O}_{\pi \pi}^{\otimes \gamma_{5}}$ & ${[0,0,1],[1,1,0]}$ \\
$\mathcal{O}_{\pi \pi}^{\otimes \gamma_{5 x / y}}, \mathcal{O}_{\pi \pi}^{\otimes \gamma_{5 z}}$ & ${[0,0,1]}$ \\
$\mathcal{O}_{\pi \pi}^{\otimes \gamma_{x t / y t}}, \mathcal{O}_{\pi \pi}^{\otimes \gamma_{z t}}$ & ${[0,0,1]}$
\end{tabular}
\caption{Operator basis on the $\approx 0.15$~fm ensemble, the single pion operators in the two-pion states have equal taste and equal and opposite momentum. We indicate the irrep splitting by separating out the operators.}
\label{table:states}
\end{table}
\vspace{-1mm}
Optimised operators with the best overlap with the state $n$ are then constructed $\chi_{n}=\left(v_{n}\right)_{i} \mathcal{O}_{i}$. Energies $E_{n}$ and overlap amplitudes $\langle 0 | J_{l} | n\rangle$ are obtained from a combined Bayesian fit to the correlation functions
\bea
\left\langle\chi_{n} \chi_{n}^{\dagger}\right\rangle&=&\sum_{n} Z_{n}^{2} e^{-E_{n} t} \\
\left\langle\chi_{n}\, J_{l}^{\dagger}\right\rangle&=&\sum_{n} Z_{n}\left\langle 0\left|J_{l}\right| n\right\rangle e^{-E_{n} t} \, . \label{gevpCorrs}
\eea
Shown in Fig.~\ref{fig:gevp} are the results from a GEVP analysis. The expected irrep splitting is observed in the two-pion states derived from the three-dimensional single-pion irreps n = 1\&2, 4\&5 are non-degenerate. We observe a reduction / enhancement of overlap with the vector current for these split irreps. The vector current correlation function is reconstructed from these states in Fig.~\ref{fig:recon}. To evaluate the efficacy of the reconstruction, we determine $a_{\mu}^{ll}$ in three different ways, (i) integrating the correlator data using Eqn.~(\ref{amuTint}), (ii) integrating using the bounding method \cite{RBC:2018dos} to address the statistical noise at large $t$, (iii) integrating the combination of the correlator data below $t^\star$ and reconstructed correlator after $t^\star$. For (ii) we chose a cut-off of $t_c  \approx 2.6$ fm, for (iii) we replace the correlator with the $N=6$ reconstruction after $t^\star \approx 1.8$. While the central values for $a_{\mu}^{ll}$ obtained from the three procedures are completely consistent, we find statistical errors of 4\% (i), 2.6\% (ii), and 0.6\% (iii), respectively, i.e.\ a 4-fold reduction of the statistical error with the reconstruction compared to the bounding method.


\section{Conclusions \& Outlook}
We provide an update on the status of our calculation of the light-quark, connected contribution to $a_{\mu}^{\mathrm{HVP}, \mathrm{LO}}$. In particular, we give a road-map to address the largest sources of uncertainty in our previous determination: statistical, continuum limit extrapolation, scale setting, and lattice corrections. We present blinded, preliminary results for the light-quark connected contribution evaluated for the intermediate window, $a_{\mu, \textrm{W}}^{ll}$. The statistical errors for this quantity are commensurate with other lattice groups. Our systematic error study is ongoing, and only after it is complete, will we unblind our final result.  We also present the results of our two-pion study, where we successfully demonstrate the technique for reconstructing the two-pion - vector correlator \cite{bruno2019exclusive, PhysRevD.101.054504} for staggered fermions. Our still preliminary results indicate a large improvement in our determination of the light-quark connected contribution to $a_{\mu}^{\mathrm{HVP}, \mathrm{LO}}$ on the $\approx 0.15$~fm ensemble. We are in the process of testing this calculation on the $\approx 0.12$~fm ensemble where the number of staggered two-pion states is significantly increased.

\section{Acknowledgements}

Fermilab is managed by Fermi Research Alliance, LLC (FRA), acting under Contract No. DE-AC02-07CH11359. This work was supported in part by U.S. DoE awards DE-SC0015655, DE-SC0010120, DE-SC0010005, NSF grant PHY17-19626 and PHY20-13064; NSF Graduate Fellowship DGE 2040434; by SRA (Spain) grant PID2019-106087GB-C21 / 10.13039/501100011033; by the Junta de Andalucía (Spain) FQM- 101, A-FQM-467-UGR18,  P18-FR-4314 (FEDER), and by URA Fermilab Visiting Scholarships in 2020 and 2021. 
Computations for this work were carried out in part on facilities of the USQCD Collaboration,
which are funded by the DOE Office of Science; using an award provided by the INCITE program; on resources of the Argonne Leadership Computing Facility,
which is a DOE Office of Science User Facility supported under contract DE-AC02-06CH11357; 
on resources of the Oak Ridge Leadership Computing Facility, which is a
DOE Office of Science User Facility supported under Contract DE-AC05-00OR22725; on resources of the National Energy Research Scientific Computing Center (NERSC),
a U.S. Department of Energy Office of Science User Facility located at LBNL, operated under Contract No. DE-AC02-05CH11231; with 
support from the ALCC program in the form of time on the
computers Summit and Theta; with resources provided by the Texas Advanced Computing Center
(TACC) at The University of Texas at Austin as part of the Frontera computing
project at the Texas Advanced Computing Center, made possible by NSF award OAC-1818253;  using an award from the XSEDE program, which is supported by NSF grant
ACI-1548562; using XSEDE Ranch through the allocation TG-MCA93S002; and using resources from the Blue Waters sustained-petascale computing project, which is supported by NSF awards OCI-0725070 and ACI-1238993, the State of Illinois, and as of December, 2019, the National Geospatial-Intelligence Agency. Blue Waters is a joint effort of the University of Illinois at Urbana-Champaign and its National Center for Supercomputing Applications.

\bibliographystyle{JHEP-notitle}
\bibliography{refs}

\end{document}